# Statistical Design of Thermal Protection System Using Physics-Informed Machine learning

Karthik R. Lyathakula[a]

[a]*Department of Mechanical and Aerospace Engineering, North Carolina State University, Raleigh, NC 27695, US*

**Abstract**

Estimating the material properties of thermal protection films is crucial for their effective design and application, particularly in high-temperature environments. This work presents a novel approach to determine the properties using uncertainty quantification simulations. We quantify uncertainty in the material properties for effective insulation by proposing a Bayesian distribution for them. Sampling from this distribution is performed using Monte Carlo simulations, which require repeatedly solving the predictive thermal model. To address the computational inefficiency of conventional numerical simulations, we develop a parametric Physics-Informed Neural Network (PINN) to solve the heat transfer problem. The proposed PINN significantly reduces computational time while maintaining accuracy, as verified against traditional numerical solutions. Additionally, we used the Sequential Monte Carlo (SMC) method to enable vectorized and parallel computations, further enhancing computational speedup. Our results demonstrate that integrating MCMC with PINN decreases computational time substantially compared to using standard numerical methods. Moreover, combining the SMC method with PINN yields multifold computational speedup, making this approach highly effective for the rapid and accurate estimation of material properties.

## 1. Introduction

Accurately assessing the thermal protection system (TPS) uncertainties is crucial for the aerospace industry. Previous studies have highlighted the importance of considering uncertainties in TPS design [1-4]. However, there is a lack of effective methodologies that combine physics-informed deep learning modeling with uncertainty quantification in TPSs. This research seeks to bridge this gap by introducing a novel approach that utilizes parametric physics-informed neural networks (PINNs) to accurately predict temperature nonlinear distributions in insulation materials while considering uncertainties. This research aims to address the need for precise and efficient calculations to evaluate the impact of PTPS uncertainties on aerospace applications. A typical thermal insulation system (TPS) is employed as a case study to investigate the impact of uncertainty on the thermal properties of materials and analyze the resulting temperature distribution along the thickness. The PINN model is trained to minimize the weighted loss which ensures that the network incorporates the underlying (soft penalties) physical principles governing the TPS behavior. By considering parametric uncertainty, the model can provide probabilistic predictions, enabling a more comprehensive assessment of TPS performance. This article presents the framework of physics informed machine learning with Markov Chain Monte Carlo method to quantify uncertainty in the thermal properties of TPS

## 2. Methods

This section discusses the methodology for design of the thermal protections system. Figure 1 shows the geometry of the thermal protection film on aero systems. The surface of the AeroSystems are subjected to high heat generation during the take-off and landing. The TPS need to be designed in such a way to maintain lower than a threshold temperature on the interface of the thermal film and the surface of the aero systems to avoid damage to the aero system.

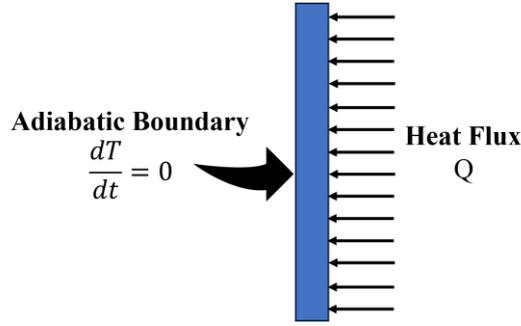

Figure 1. Geometry of the thermal protection film

The equations that represent the heat transfer in the thermal film is given by

$$\frac{\partial T}{\partial t} = \frac{k}{\rho c_p}\frac{\partial^2 T}{\partial x^2} \tag{1}$$

and subjected to adiabatic condition at the interface between film and the surface. The outer surface is subjected to constant heat flux for the flight duration. The material properties density ($\rho$), thermal conductivity ($k$), and specific heat capacity ($c_p$) are uncertain. The goal is to identify the material properties uncertainties of the thermal protection film to ensure that the interface temperature between the film and the surface remains below a specified threshold. However, uncertainties in the thermal properties make it challenging to maintain this temperature. In this study, these uncertainties are quantified, and samples of the thermal properties are generated. The uncertainties in the material properties, $\boldsymbol{q} = (\rho, k, c_p)$, is defined by Bayes theorem

$$\pi(\boldsymbol{q}|T) = \frac{\pi(T|\boldsymbol{q})\pi(\boldsymbol{q})}{\int_q \pi(T|\boldsymbol{q})\pi(\boldsymbol{q})d\boldsymbol{q}} \tag{2}$$

where $\pi(\boldsymbol{q}|T)$ is the posterior distribution, $\pi(T|\boldsymbol{q})$ is the likelihood, $\pi(\boldsymbol{q})$ is prior distribution. The uncertainty in the parameters is quantified by generating the samples from the above distribution using Monte Carlo methods [5]. In the Eq. 2, the prior is assumed to be normal distribution. Assuming the $\pi(T|\boldsymbol{q})$ is normal distribution with mean $\mu$ and standard deviation $\sigma$

$$\pi(T|\boldsymbol{q}) \sim N(\mu, \sigma) \tag{3}$$

Using Eq.4 and ignoring the normalization factor, Eq. 2 can be simplified as

$$\pi(\boldsymbol{q}|T) \alpha\, \pi(T|\boldsymbol{q}) \tag{4}$$

Some of the generated parameters, however, may result in interface temperatures exceeding the threshold due to these uncertainties. To account for this, a reliability factor is introduced to evaluate the performance of the thermal protection system (TPS) under uncertain conditions. The reliability of the TPS is defined by [3] following equation

$$R = \frac{N(T_{max} < T_{th})}{N} \times 100 \% \qquad (5)$$

where $N$ is the number of possible parameter samples generated from the parametric posterior, $N(T_{max} < T_{th})$ is the number of samples the temperature on the interface is below the threshold. For given $T_{th}$, $R$ and $\sigma$ of the likelihood function, the likelihood function distribution is given by

$$\pi(T|\boldsymbol{q}) \alpha \, N(\mu = T_{th} - z_p(R)\sigma, \sigma) \qquad (6)$$

The uncertainty in the parameters are estimated by generating samples from Eq. (6)

For generating samples, the heat transfer equations (1) need to be solved. In the statistical design of Thermal Protection Systems (TPS), there will be multiple combinations of thermal property samples—density (ρ), thermal conductivity (k), and specific heat capacity ($c_p$)—that satisfy a specified interface temperature. Markov Chain Monte Carlo (MCMC) methods are utilized to produce these thermal property samples by generating samples from (5), which require multiple heat transfer simulations. However, traditional numerical approaches are computationally expensive and significantly increase computational time. To address this challenge and achieve efficient solutions, Physics-Informed Neural Networks (PINNs) are employed [3, 6-8], offering a faster and more computationally efficient alternative for solving complex numerical problems.

*Figure 2* shows the overview of PINN for simulating the behavior of the TPS. The residue of the heat transfer equation is defined as the loss function and the weight of the neural network is obtained by minimizing the loss function. Automatic differentiation used to accurately calculate each term in the differential equation [3, 9]

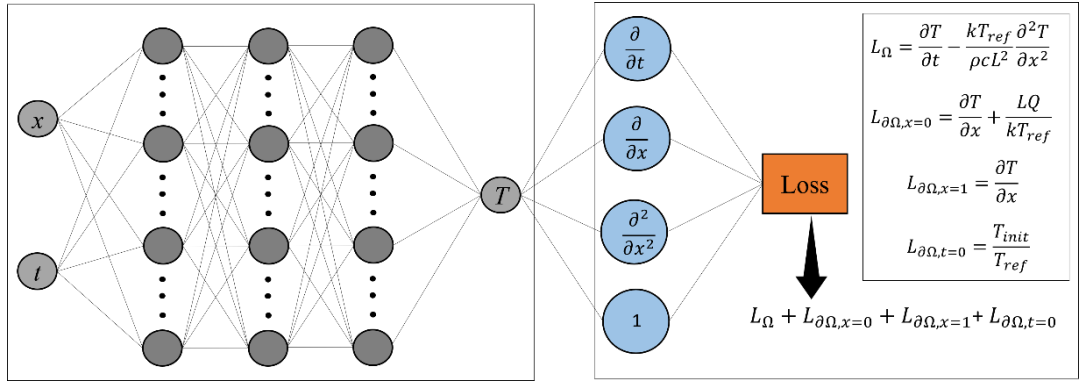

Figure 2. Physics informed neural network for solving the thermal behavior of TPS

## 3. Results

This section presents the results, beginning with the PINN model, followed by statistical simulations. Figure 3 illustrates the temperature distribution results obtained from the thermal simulation using PINN, compared to numerical and reference simulations. The material properties for the thermal simulation used from reference [3] . The maximum error in the PINN simulation across the domain is 27°C. The objective is to estimate the temperature at t = 150 seconds and at the interface. At the point of interest, the absolute error in the PINN prediction is less than 2°C. However, once trained the PINN reduces the computational time exponentially using vectorization

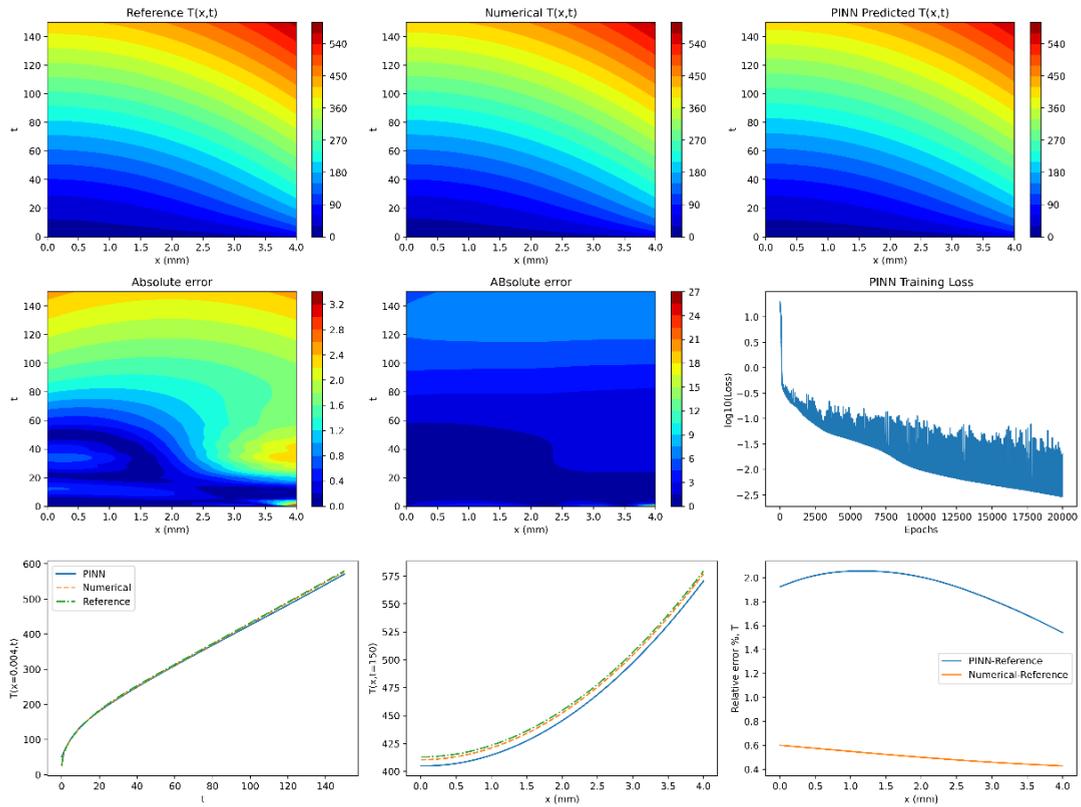

Figure 3 Comparison of temperature distribution from PINN compared to numerical simulation and reference

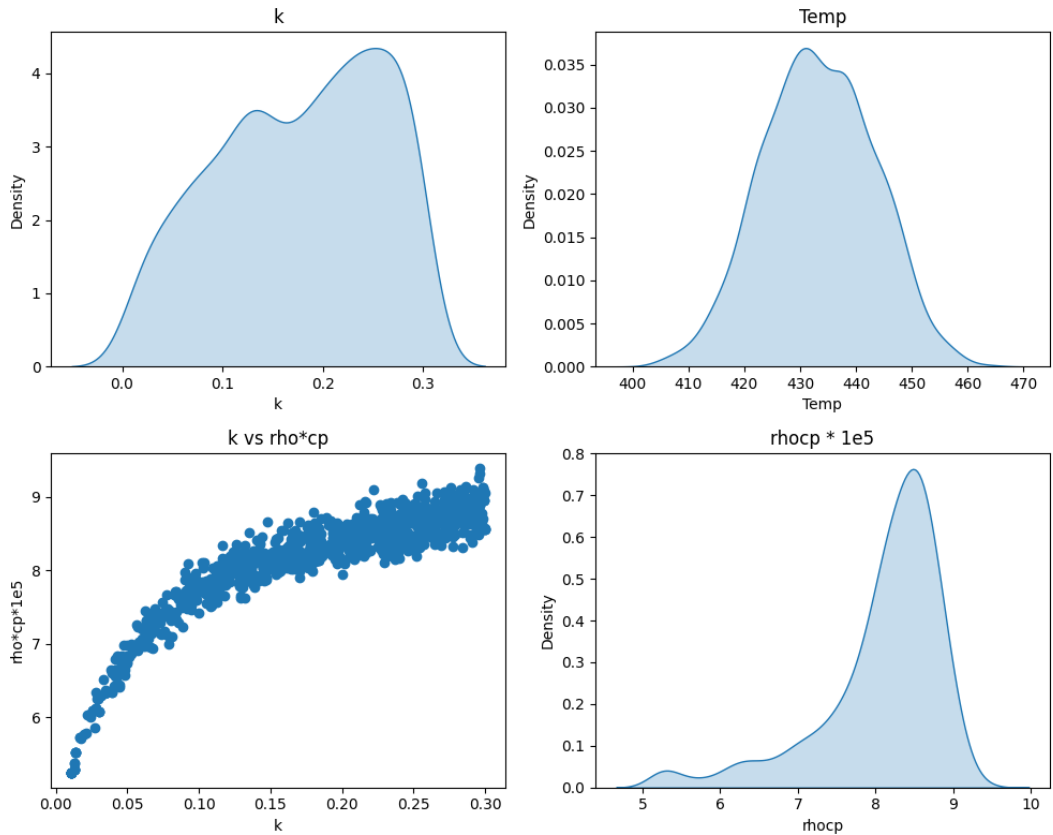

Figure 4 Statistical samples of the thermal parameters maintain the temperature at the film interface to below 450 C with reliability of 95%.

Next using the PINN the statistical simulation are conducted by generating samples from Normal distribution presented in Eq. (6) using Markov chain Monte Carlo method [5, 10, 11] . Figure 4 shows the statistical samples of the thermal parameters maintaining the temperature at the film interface to below 450 C with reliability of 95%.

References

[1] Uyanna, O., and Najafi, H., "Thermal protection systems for space vehicles: A review on technology development, current challenges and future prospects," *Acta Astronautica,* Vol. 176, 2020, pp. 341–356.

[2] Nakamura, T., and Fujii, K., "Probabilistic transient thermal analysis of an atmospheric reentry vehicle structure," *Aerospace Science and Technology,* Vol. 10, No. 4, 2006, pp. 346–354.

[3] Zhang, R., Xu, N., Zhang, K., "A Parametric Physics-Informed Deep Learning Method for Probabilistic Design of Thermal Protection Systems," *Energies,* Vol. 16, No. 9, 2023, pp. 3820.

[4] Dong, Y., Wang, E., You, Y., "Thermal protection system and thermal management for combined-cycle engine: Review and prospects," *Energies,* Vol. 12, No. 2, 2019, pp. 240.

[5] Smith, R.C., "Uncertainty quantification: theory, implementation, and applications," SIAM, Philadelphia, PA, 2014, 2014,

[6] Haghighat, E., Raissi, M., Moure, A., "A physics-informed deep learning framework for inversion and surrogate modeling in solid mechanics," *Computer Methods in Applied Mechanics and Engineering,* Vol. 379, 2021, pp. 113741.

[7] Raissi, M., Perdikaris, P., and Karniadakis, G.E., "Physics-informed neural networks: A deep learning framework for solving forward and inverse problems involving nonlinear partial differential equations," *Journal of Computational Physics,* Vol. 378, 2019, pp. 686–707.

[8] Lyathakula, K.R., Cesmeci, S., DeMond, M., "Physics-Informed Deep Learning-Based Proof-of-Concept Study of a Novel Elastohydrodynamic Seal for Supercritical CO2 Turbomachinery," *Journal of Energy Resources Technology,* Vol. 145, No. 12, 2023,

[9] Baydin, A.G., Pearlmutter, B.A., Radul, A.A., "Automatic differentiation in machine learning: a survey," *Journal of Machine Learning Research,* Vol. 18, No. 153, 2018, pp. 1–43.

[10] Lyathakula, K.R., and Yuan, F.G., "A Probabilistic Fatigue Life Prediction for Adhesively Bonded Joints via ANNs-based Hybrid Model," *International Journal of Fatigue,* 2021, pp. 106352.

[11] Leser, P.E., Hochhalter, J.D., Warner, J.E., "Sequential Monte Carlo: Enabling Real-time and High-fidelity Prognostics," *Annual Conference of the PHM Society,* Vol. 10, 2018,